\documentclass[12pt]{article} 
\usepackage{amssymb} 
 
\usepackage{graphics} 
\usepackage{amsmath} 
\newcommand{\gsim}{\mbox{ \raisebox{-4pt}{${\stackrel{\textstyle >}{\sim}}$} }}

\newcommand{\tb}{\ensuremath{\tan\beta}}  
\newcommand{\stackm}{\stackrel{\scriptstyle <}{{ }_{\sim}}}  
\newcommand{\stackM}{\stackrel{\scriptstyle >}{{ }_{\sim}}}

\hbox{   
\vrule height0pt width5in   
\vbox{\hbox{\rm   
}\break   
\hbox{UAB-FT-491 \hfill}   
\hbox{KA-TP-22-2000 \hfill}   
\hbox{hep-ph/0011091} 
\hrule height.1cm width0pt}} \vspace{3mm}   
\begin{document}

\vskip1.3cm 
 
\begin{center} 
{\Large Loop Induced Flavor Changing Neutral Decays of the Top Quark in a 
General Two-Higgs-Doublet Model} 
 
\vskip8mm{\large Santi B\'{e}jar}$^{a}${\large , Jaume Guasch}$^{b}${\large ,  

Joan Sol\`{a}}$^{a}$ \vskip5mm 
 
\medskip 
 
$^{a}$\textsl{Grup de F{\'\i}sica Te\`{o}rica and Institut de F{\'\i}sica 
d'Altes Energies, } 
 
\textsl{Universitat Aut\`{o}noma de Barcelona, E-08193, Bellaterra, 
Barcelona, Catalonia, Spain} 
 
$^{b}$\textsl{Institut f\"{u}r Theoretische Physik, 
Universit{\"a}t Karlsruhe, Kaiserstra{\ss}e 12, D-76128 Karlsruhe,
Germany.} 
\end{center} 
 
\vspace{1.3cm} 
 
\begin{center} 
\textbf{ABSTRACT} 
\end{center} 
 
\begin{quotation} 
Decays of the top quark induced by flavor changing neutral currents
(FCNC) are known to be extremely rare events within the Standard
Model. This is so not only for the decay modes into gauge bosons, but
most notably in the case of the Higgs channels, e.g. $t\rightarrow
H_{SM}+c$, with a branching fraction of $10^{-13}$ at most. Therefore,
detection of FCNC top quark decays in a future high-energy, and
high-luminosity, machine like the LHC or the LC would be an indisputable
signal of new physics. In this paper we show that within the simplest
extension of the SM, namely the general two-Higgs-doublet model, the
FCNC top quark decays into Higgs bosons,
$t\rightarrow(h^{0},H^{0},A^{0})+c$, can be the most favored FCNC modes
-- comparable or even more efficient than the gluon channel
$t\rightarrow g+c$.  In both cases the optimal results are obtained for
Type~II models. However, only the Higgs channels can have rates reaching
the detectable level ($10^{-5}$), with a maximum of order $10^{-4}$
which is compatible with the charged Higgs bounds from radiative B-meson
decays. We compare with the previous results obtained in the Higgs
sector of the MSSM.
\end{quotation} 
 
\newpage 
 
\section{Introduction} 
 
%%%%%%%%%%%%%%%%%%%%%%%%%%%%%%%%%%%%%%%%%%%%%%%%%%%%%%%%%   
%  
%  
%  
%  
%  
%  
 
In the near and middle future, with the upgrades of the Tevatron (Run II, 
TeV33), the advent of the LHC, and the construction of an $e^{+}e^{-}$ 
linear collider (LC), new results on top quark physics\,\cite{TopPhys},
and possibly also on  
Higgs physics, will be obtained that may be extremely helpful to complement 
the precious information already collected at LEP 100 and 200 from Z and W 
physics. Both types of machines, the hadron colliders and the LC will work 
at high luminosities and will produce large amounts of top quarks. In the 
LHC, for example, the production of top quark pairs will be $\sigma(t% 
\overline{t})=800\;pb$ -- {roughly two orders of magnitude larger than in 
the Tevatron Run II}. In the so-called low-luminosity phase ($% 
10^{33}\,cm^{-2}s^{-1}$) of the LHC one expects about three $t\,\bar{t}$% 
-pair per second (ten million $t\,\bar{t}$-pairs per year!)~\cite{Gianotti}. 
And this number will be augmented by one order of magnitude in the 
high-luminosity phase ($10^{34}\,cm^{-2}s^{-1}$). As for a future LC running 
at e.g. $\sqrt{s}=500\;GeV$, one has a smaller cross-section $\sigma(t\bar{t}% 
)=650\;fb$ but a higher luminosity factor ranging from $5\times10^{33}% 
\,cm^{-2}s^{-1}$ to $5\times10^{34}\,cm^{-2}s^{-1}$ and of course a much 
cleaner environment\thinspace\cite{Miller}. With datasets from LHC and LC 
increasing to several $100\,fb^{-1}/$year in the high-luminosity phase, one 
should be able to pile up an enormous wealth of statistics on top quark 
decays. Therefore, not surprisingly, these machines should be very useful to 
analyze rare decays of the top quark, viz. decays whose branching fractions 
are so small ($\stackm10^{-5}$) that they could not be seen unless the 
number of collected top decays is very large. 
 
The reason for the interest in these decays is at least twofold. First, the 
typical branching ratios for the rare top quark decays predicted within the 
Standard Model (SM) are so small that the observation of a single event of 
this kind should be ``instant evidence'', so to speak, of new physics; and 
second, due to its large mass ($m_{t}=174.3\pm5.1\,GeV$~\cite{PDB2000}), the 
top quark could play a momentous role in the search for Higgs physics beyond 
the SM. While this has been shown to be the case for the top quark decay 
modes into charged Higgs bosons, both in the Minimal Supersymmetric Standard 
Model (MSSM) and in a general two-Higgs-doublet model (2HDM)~\cite 
{CGGJS,CGHS}\footnote{% 
For a recent review of the main features of loop-induced supersymmetric 
effects on top quark production an decay, see e.g. Ref.~\cite{DESY}.}, we 
expect that a similar situation could apply for top quark decays into non-SM 
neutral Higgs bosons. Notice that the latter are rare top quark events of a 
particularly important kind: they are decays of the top quark mediated by flavor changing neutral currents (FCNC). 
 
At the tree-level there are no FCNC processes in the SM, and at one-loop 
they are induced by charged-current interactions, which are GIM-suppressed\,\cite{GIM}. 
In particular, FCNC decays of the top quark into gauge bosons ($t\rightarrow 
c\,V$;$\;V\equiv\gamma,Z,g$) are very unlikely. For the present narrow range 
of masses for the top quark, they yield maximum branching ratios of $% 
\sim5\times10^{-13}$ for the photon, slightly above $1\times10^{-13}$ for 
the $Z$-boson, and $\sim4\times10^{-11}$ for the gluon channel\thinspace  
\cite{GEilam}. These are much smaller than the FCNC rates of a typical 
low-energy meson decay, e.g. $B(b\rightarrow s\,\gamma)\sim10^{-4}$. And the 
reason is simple: for FCNC top quark decays in the SM, the loop amplitudes 
are controlled by down-type quarks, mainly by the bottom quark. Therefore, 
the scale of the loop amplitudes is set by $m_{b}^{2}$ and the partial 
widths are of order  
\begin{equation} 
\Gamma(t\rightarrow 
V\,c)\sim|V_{tb}^{\ast}V_{bc}|^{2}\alpha\,G_{F}^{2}\,m_{t}\,m_{b}^{4}\,F% 
\sim|V_{bc}|^{2}\alpha_{em}^2 \alpha \,m_{t}\left( \frac{m_{b}}{M_{W}}\right) ^{4}\,F, 
\label{GammaFCNC} 
\end{equation} 
where $\alpha$ is $\alpha_{em}$ for $V=\gamma,Z$ and $\alpha_{s}$ for $V=g$. 
The factor $F\sim(1-m_{V}^{2}/m_{t}^{2})^{2}$ results, upon neglecting $m_{c}$, 
from phase space and polarization sums. Notice that the 
dimensionless fourth power mass ratio, in parenthesis in eq.~(\ref{GammaFCNC}% 
), stems from the GIM mechanism and is responsible for the ultralarge 
suppression beyond naive expectations based on pure dimensional analysis, 
power counting and CKM matrix elements. From that simple formula, the approximate orders of 
magnitude mentioned above ensue immediately. 
 
Even more dramatic is the situation with the top quark decay into the SM 
Higgs boson, $t\rightarrow c\,H_{SM}$, which has recently been recognized to 
be much more disfavored than originally claimed \thinspace\cite{GEilam}: $% 
BR(t\rightarrow c\,H_{SM})\sim10^{-13}-10^{-15}$ $(m_{t}=175\,GeV;\;M_{Z}% 
\leq M_{H}\leq2\;M_{W})$\thinspace\cite{Mele}. This extremely tiny rate is 
far out of the range to be covered by any presently conceivable high 
luminosity machine. On the other hand, the highest FCNC top quark rate in 
the SM, namely that of the gluon channel $t\rightarrow c\,g$, is still $6$ 
orders of magnitude below the feasible experimental possibilities at the 
LHC. All in all the detection of FCNC decays of the top quark at visible 
levels (viz. $\,BR(t\rightarrow c\,X)>10^{-5}$) by the high luminosity 
colliders round the corner (especially LHC and LC) seems doomed to failure 
in the absence of new physics. Thus the possibility of large enhancements of 
some FCNC channels up to the visible threshold, particularly within the 
context of the general 2HDM and the 
MSSM, should be very welcome. Unfortunately, although the FCNC decay modes 
into electroweak gauge bosons $V_{ew}=W,Z$ may be enhanced a few orders of 
magnitude, it proves to be insufficient to raise the meager SM rates 
mentioned before up to detectable limits, and this is true both in the 2HDM 
-- where $BR(t\rightarrow V_{ew}\,c)<10^{-6}$~\cite{GEilam} -- and in the 
MSSM -- where $BR(t\rightarrow V_{ew}\,c)<10^{-7}$ except in highly unlikely 
regions of the MSSM parameter space~\cite{Lopez}\footnote{% 
Namely, regions in which there are wave-function renormalization thresholds 
due to (extremely fortuitous!) sharp coincidences between the sum of the 
sparticle masses involved in the self-energy loops and the top quark mass. 
See e.g. Ref.~\cite{GJHS} for similar situations already in the conventional  
$t\rightarrow W\,b$ decay within the MSSM. In our opinion these narrow 
regions should not be taken too seriously.}. In this respect it is a lucky 
fact that these bad news need not to apply to the gluon channel, which could 
be barely visible ($BR(t\rightarrow g\,c)\stackm10^{-5}$) both in the
MSSM~\cite{Divitiis,NP} and in the general 2HDM~\cite{GEilam}. But, most  
significant of all, they may not apply to the non-SM Higgs boson channels $% 
t\rightarrow(h^{0},H^{0},A^{0})+c$ either. As we shall show in the sequel, 
these Higgs decay channels of the top quark could lie above the visible 
threshold for a parameter choice made in perfectly sound regions of 
parameter space! 
 
While a systematic discussion of these ``gifted'' Higgs channels
was made in Ref.~\cite{NP} for the MSSM 
case and in other models\footnote{Preliminary SUSY analysis of the Higgs channels
are given in \cite 
{Hewett}, but they assume the MSSM Higgs mass relations at the 
tree-level. Therefore these are particular cases of the general MSSM 
approach given in~\cite{NP}. Studies beyond the MSSM (e.g. including 
R-parity violation) and also in quite different contexts from the present 
one are available in the 
literature, see e.g.\thinspace \cite{FCNCSM}.}, 
to the best of our knowledge there is no similar study in the general 
2HDM. And we believe that this study is 
necessary, not only to assess what are the chances to see traces of new 
(renormalizable) physics in the new colliders round the corner but also to 
clear up the nature of the virtual effects; in particular to disentangle 
whether the origin of the hypothetically detected FCNC decays of the top 
quark is ultimately triggered by SUSY or by some alternative, more generic, 
renormalizable extension of the SM such as the 2HDM or generalizations 
thereof. Of course the alleged signs of new physics could be searched for 
directly through particle tagging, if the new particles were not too heavy. 
However, even if accessible, the corresponding signatures could be far from 
transparent. In contrast, the indirect approach based on the FCNC processes 
has the advantage that one deals all the time with the dynamics of the top 
quark. Thus by studying potentially new features beyond the well-known SM 
properties of this quark one can hopefully uncover the existence of the 
underlying new interactions. 
 
\section{Relevant fields and interactions in the 2HDM} 
 
%%%%%%%%%%%%%%%%%%%%%%%%%%%%%%%%%%%%%%%%%%%%%%%%%%%%%%%   
%  
%  
%  
%  
%  
%  

We will mainly focus our interest on the loop induced FCNC decays  
\begin{equation} 
t\rightarrow c\;h\;\;\;(h=h^{0},H^{0},A^{0})\,,  \label{Higgschannels} 
\end{equation} 
in which any of the three possible neutral Higgs bosons from a general 2HDM 
can be in the final state. However, as a reference we shall compare 
throughout our analysis the Higgs channels with the more conventional gluon 
channel  
\begin{equation} 
t\rightarrow c\;g\,.  \label{gchannel} 
\end{equation} 
\begin{figure}[t] 
\begin{center} 
%\resizebox{\textwidth}{!}{
\includegraphics{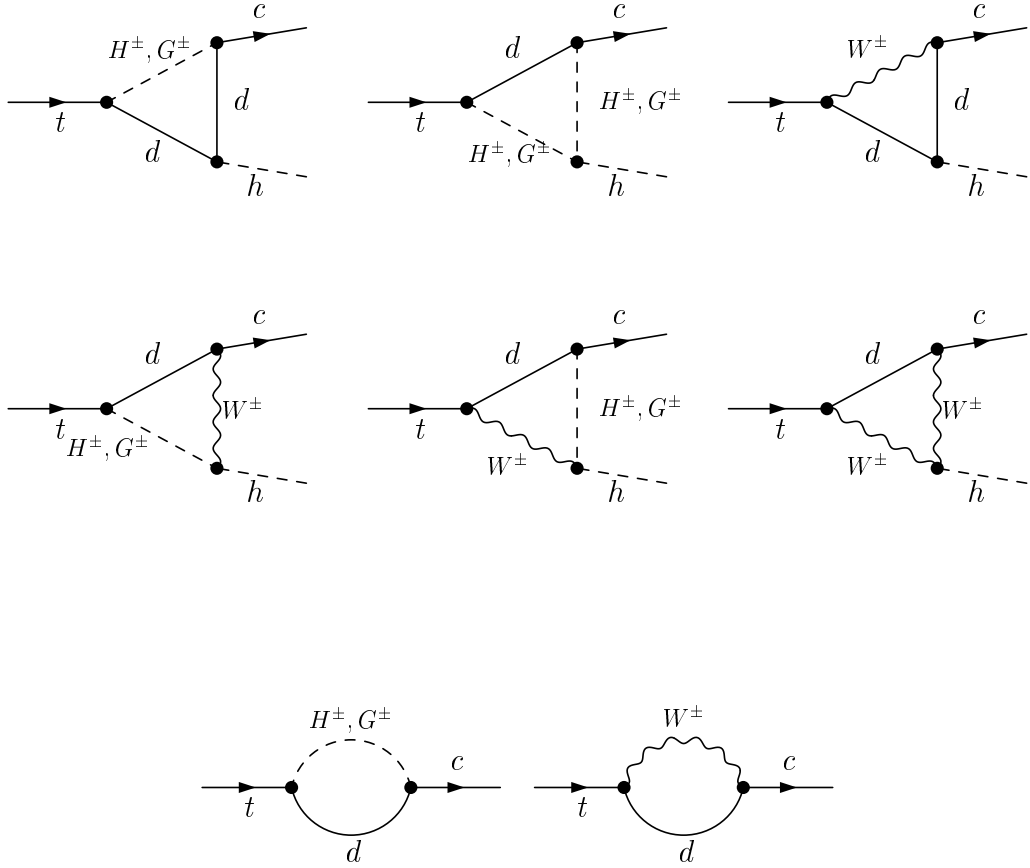}%} 
\end{center} 
\caption{One-loop vertex diagrams contributing to the FCNC top quark decays (% 
\ref{Higgschannels}). Shown are the vertices and mixed self-energies with 
all possible contributions from the SM fields and the Higgs bosons from the 
general 2HDM. The Goldstone boson contributions are computed in the Feynman 
gauge.} 
\label{fig:1} 
\end{figure}% 
Although other quarks could participate in the final state of these 
processes, their contribution is obviously negligible -- because it is 
further CKM-suppressed. The lowest order diagrams entering these decays are 
one-loop diagrams in which Higgs, quarks, gauge and Goldstone bosons -- in 
the Feynman gauge -- circulate around. While the diagrams for the decays (% 
\ref{Higgschannels}) are depicted in Fig.~\ref{fig:1}, the ones for the 
decay (\ref{gchannel}) are not shown~\cite{GEilam}. Here we follow the 
standard notation~\cite{Hunter}, namely $h^{0},H^{0}$ are CP-even Higgs 
bosons (where by convention $m_{h^{0}}<m_{H^{0}}$) and $A^{0}$ is a CP-odd 
one (often called a ``pseudoscalar''). As it is well-known, when the quark 
mass matrices are diagonalized in non-minimal extensions of the Higgs sector 
of the SM, the Yukawa couplings do not in general become simultaneously 
diagonalized, so that one would expect Higgs mediated FCNC's at the 
tree-level. These are of course unwanted, since they would lead to large 
FCNC processes in light quark phenomenology, which are stringently 
restricted by experiment. Apart from letting the Higgs masses to acquire 
very large values, one has two additional, more elegant, canonical choices 
to get rid of them. In Type~I 2HDM (also denoted 2HDM~I) one Higgs doublet, $% 
\Phi_{1}$, does not couple to fermions at all and the other Higgs doublet, $% 
\Phi_{2}$, couples to fermions in the same manner as in the SM. In contrast, 
in Type~II 2HDM (also denoted 2HDM~II) one Higgs doublet, $\Phi_{1}$, 
couples to down quarks (but not to up quarks) while $\Phi_{2}$ does the 
other way around. Such a coupling pattern is automatically realized in the 
framework of supersymmetry (SUSY), in particular in the MSSM, but it can 
also be arranged if we impose a discrete symmetry, e.g. $\Phi 
_{1}\rightarrow-\Phi_{1}$ and $\Phi_{2}\rightarrow+\Phi_{2}$ (or vice versa) 
plus a suitable transformation for the right-handed quark fields. We shall 
not worry here on the ultimate theoretical origin of this ad hoc symmetry, 
but we will accept it as a guiding principle. As mentioned above, the 
SUSY case has recently been investigated 
in Ref.~\cite{NP}, so we wish to concentrate here on Type~I and Type~II models of a sufficiently generic nature, to wit, those which are 
characterized by the following set of free parameters:  
\begin{equation} 
(m_{h^{0}},m_{H^{0}},m_{A^{0}},m_{H^{\pm}},\tan\alpha,\tan\beta )\,, 
\label{freeparam} 
\end{equation} 
where $m_{H^{\pm}}$ is the mass of the charged Higgs companions $H^{\pm}$, $% 
\tan\alpha$ defines the mixing angle $\alpha$ in the diagonalization of the 
CP-even sector, and $\tan\beta$ gives the mixing angle $\beta$ in the CP-odd 
sector. The latter is a key parameter in our analysis. It is given by the 
quotient of the vacuum expectation values (VEV's) of the two Higgs doublets $% 
\Phi_{2,1}$, viz. $\tan\beta=v_{2}/v_{1}$, where the parameter sum $% 
v^{2}\equiv v_{1}^{2}+v_{2}^{2}$ is fixed by the $W$ mass: $% 
M_{W}^{2}=(1/2)\,g^{2}\,v^{2}$ ($g$ is the weak $SU(2)$ gauge coupling) or, 
equivalently, by the Fermi constant: $G_{F}=1/(2\sqrt{2})\,v^{2}$. It is 
well-known~\cite{Hunter} that the most general 2HDM Higgs potential subject 
to hermiticity, $SU(2)\times U(1)$ gauge invariance and a discrete symmetry 
of the sort mentioned above involves six scalar operators with six free 
(real) coefficients $\lambda_{i}\,(i=1,2,\ldots,6)$ and the two VEV's% 
\footnote{% 
By imposing the discrete symmetry one is able, in principle, to get rid of 
two additional quartic and one bilinear operators in the Higgs potential. 
These quartics are not shown in~\cite{Hunter}, but the most general 2HDM 
potential could contain all of these additional terms~\cite{TASI}. The 
bilinear ones are eventually kept in most cases as one usually makes 
allowance for the discrete symmetry to be only (softly) violated by the 
dimension two operators. The resulting model is still a minimal setup 
compatible with the absence of Higgs-mediated dangerous FCNC.}. We will 
furthermore assume that $\lambda _{5}=\lambda_{6}$ in the general 2HDM Higgs 
potential~\cite{GunionH}. This allows to remove the CP phase in the 
potential by redefining the phase of one of the doublets. In this way one 
can choose the VEV's of $\Phi_{1,2}$ real and positive. The alternative set~(% 
\ref{freeparam}) is just a (more physical) reformulation of this fact after 
diagonalization of the mass matrices and imposing the aforementioned set of 
constraints.

As stated above, the two canonical types of 2HDM's only differ in the 
couplings to fermions but they share Feynman rules generally different from 
the corresponding ones in the MSSM~\cite{Hunter}. The necessary Feynman 
rules to compute the diagrams in Fig.~\ref{fig:1} are mostly well-known. For 
instance, the interactions between Higgs and Goldstone bosons and the 
interactions between Goldstone bosons and gauge bosons are the same as in 
the MSSM~\cite{Hunter}. Similarly for the interactions among Goldstone 
bosons and gauge bosons, and Goldstone bosons and fermions. However, the 
2HDM Feynman rules for the charged and neutral Higgs interactions with 
fermions and also for the trilinear self-couplings of Higgs bosons may 
drastically deviate from the MSSM. The charged Higgs interactions with 
fermions are encoded in the following interaction Lagrangian  
\begin{equation} 
\mathcal{L}_{Htb}^{(j)}=\frac{gV_{tb}}{\sqrt{2}\,M_{W}}\,H^{-}\overline{b}\,% 
\left[ m_{t}\cot \beta \,P_{R}+m_{b}\,a_{j}\,P_{L}\right] \,t+h.c. 
\label{Htb} 
\end{equation} 
where here, and hereafter, we use third-quark-family notation as a generic one;
$V_{tb}$ is
the corresponding CKM matrix element, $P_{L,R}=(1/2)(1\mp \gamma _{5})$ 
are the chiral projection operators on left- and right-handed fermions, $% 
j=I,II$ runs over Type~I and Type~II 2HDM's, and we have introduced a 
parameter $a_{j}$ such that $a_{I}=-\cot \beta $ and $a_{II}=+\tan \beta $. 
For the neutral Higgs interactions, \ the necessary pieces of the Lagrangian 
(see Fig.~\ref{fig:1}) can be written in the compact form  
\begin{eqnarray} 
\mathcal{L}_{hqq}^{(j)} &=&\frac{-g\,m_{b}}{2\,M_{W}\,\left\{  
\begin{array}{c} 
\sin \beta \\  
\cos \beta 
\end{array} 
\right\} }\,\overline{b}\left[ h^{0}\,\left\{  
\begin{array}{c} 
\cos \alpha \\  
-\sin \alpha 
\end{array} 
\right\} +H^{0}\,\left\{  
\begin{array}{c} 
\sin \alpha \\  
\cos \alpha 
\end{array} 
\right\} \right] \,b+\frac{i\,g\,m_{b}\,a_{j}}{2\,M_{W}}\,\overline{b}% 
\,\gamma _{5}\,b\,A^{0}  \notag \\ 
&+&\frac{-g\,m_{t}}{2\,M_{W}\,\sin \beta }\,\overline{t}\left[ h^{0}\,\cos 
\alpha +H^{0}\,\sin \alpha \right] \,t+\frac{i\,g\,m_{t}}{2\,M_{W}\tb}\,% 
\overline{t}\,\gamma _{5}\,t\,A^{0}\,\,,  \label{Hff} 
\end{eqnarray} 
\begin{table}[t] 
\begin{center} 
\includegraphics{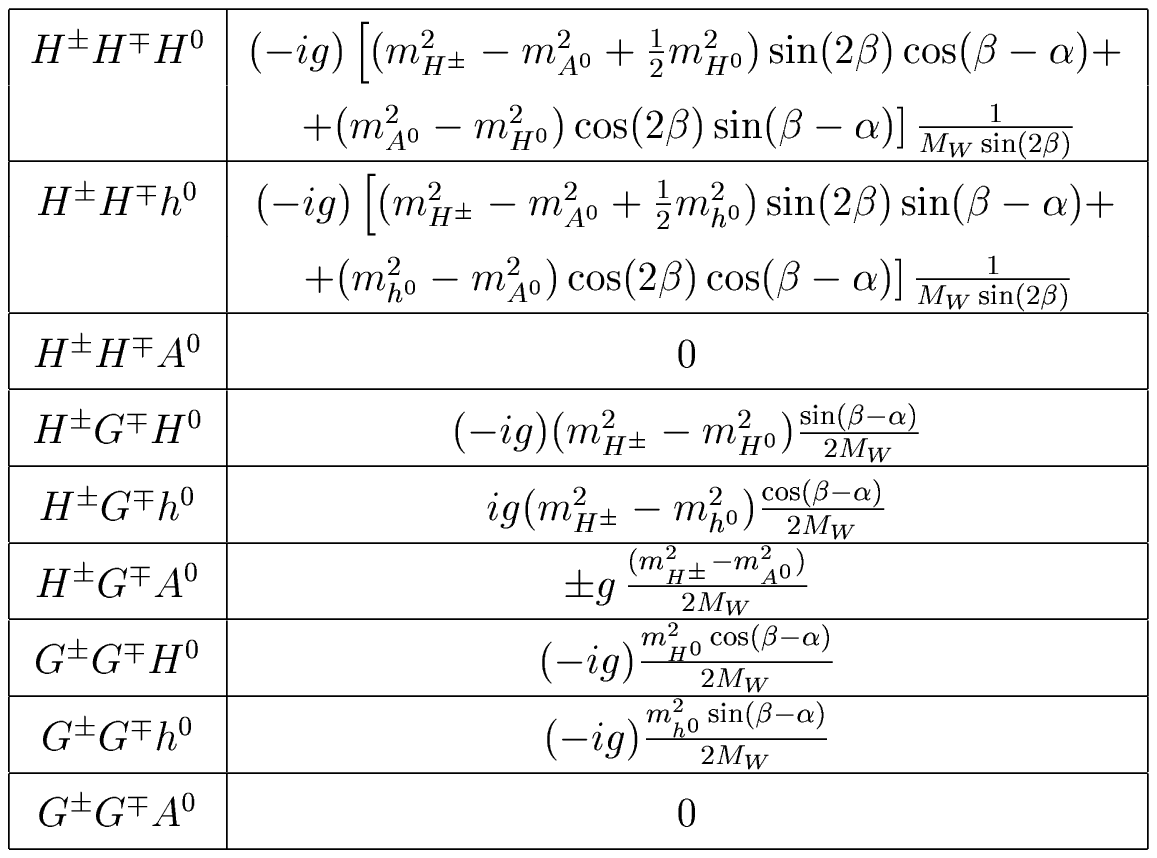}
\end{center} 
\caption{Feynman rules for the trilinear couplings involving the Higgs 
self-interactions and the Higgs and Goldstone boson vertices in the Feynman 
gauge, with all momenta pointing inward. These rules are common to both Type~I 
and Type~II 2HDM under the conditions explained in the text. We have singled out 
some null entries associated to CP violation. The null vertices imply the 
corresponding deletion of some vertex diagrams in Fig.~\ref{fig:1}.} 
\label{tab:1} 
\end{table}% 
where the upper row is for $j=I$ and the down row is for $j=II$. 
As far as 
it goes to the 2HDM Feynman rules for the trilinear couplings among Higgs bosons, 
and Higgses with Goldstone bosons, they are summarized in Table~\ref{tab:1}, 
and are valid for Type~I and Type~II models. Let us recall that Type~II 
models are specially important in that the Higgs sector of the MSSM is 
precisely of this sort. Had we not imposed the restriction $\lambda 
_{5}=\lambda _{6}$, then the trilinear rules would be explicitly dependent 
on the $\lambda _{5}$ parameter. However the numerical analysis that we 
perform in the next section does not depend in any essential way on this 
simplification. In essence we have just traded $\lambda _{5}$ for $% 
m_{A^{0}}^{2}$ in these rules and so by varying with respect to $m_{A^{0}}$ 
we do explore most of the quantitative potential of the general 2HDM. In the 
MSSM case the condition $\lambda _{5}=\lambda _{6}$ is automatic by the 
underlying supersymmetry, and the values of these couplings are determined 
by the $SU(2)\times U(1)$ electroweak gauge couplings. But of course we 
treat here the parameters of Type~II models in a way not restricted by SUSY 
prejudices. On the other hand there is no need to depart arbitrarily from 
the SUSY frame, as it can be useful for a better comparison. Be as it may, 
our Type~II Higgs model is still sufficiently general that it cannot be 
considered as the limit of the MSSM when all the sparticle masses are 
decoupled. Both in the generic 2HDM~II and in the MSSM, the Feynman rules 
for the lightest CP-even Higgs, $h^{0}$, go over to the SM Higgs boson ones 
in the limit $\sin (\beta -\alpha )\rightarrow 1$. In the particular case of 
the MSSM, but \emph{not} in a general 2HDM~II, this limit is equivalent to $% 
m_{A^{0}}\rightarrow \infty $. Moreover, in the MSSM one 
has\footnote{See \cite{Hollik} and references therein for the status 
of the Higgs mass calculations in the MSSM.} $m_{h^{0}}\stackm%  
135\,GeV$ 
whereas in the general Type~II model there is no   
upper bound on $m_{h^{0}}$, and by the same token the corresponding lower 
bound is considerably less stringent (see below). 
 
Since we shall perform our calculation in the on-shell scheme, we understand 
that the physical inputs are given by the electromagnetic coupling and the 
physical masses of all the particles:  
\begin{equation} 
(e,M_{W},M_{Z},m_{h^{0}},m_{H^{0}},m_{A^{0}},m_{H^{\pm}},m_{f})\,. 
\label{onshell} 
\end{equation} 
The remaining parameters, except the Higgs mixing angles, are understood to 
be given in terms of the latter, e.g. the $SU(2)$ gauge coupling appearing 
in the previous formulae and in Table~\ref{tab:1} is given by $g=e/s_{w}$ , 
where the sinus of the weak mixing angle is defined through $% 
s_{w}^{2}=1-M_{W}^{2}/M_{Z}^{2}$. It should be clear that, as there are no 
tree-level FCNC decays of the top quark, there is no need to introduce 
counterterms for the physical inputs in this calculation. In fact, the 
calculation is carried out in lowest order (``tree level'') with respect to 
the effective $tch$ and $tcg$ couplings and so the sum of all the one-loop 
diagrams (as well as of certain subsets of them) should be finite in a 
renormalizable theory, and indeed it is. 
 
\section{Numerical analysis} 
 
From the previous interaction Lagrangians and Feynman rules it is now 
straightforward to compute the loop induced FCNC rates for the decays (\ref 
{Higgschannels}) and (\ref{gchannel}). We shall refrain from listing the 
lengthy analytical formulae as the computation is similar to the one 
reported in great detail in Ref.~\cite{NP}. Therefore, we will limit 
ourselves to exhibit the final numerical results. The fiducial ratio on 
which we will apply our numerical computation is the following:  
\begin{equation} 
B^{j}(t\rightarrow h+c)=\frac{\Gamma^{j}(t\rightarrow h+c)}{\Gamma 
(t\rightarrow W^{+}+b)+\Gamma^{j}(t\rightarrow H^{+}+b)}\,\,, 
\label{fiducialH} 
\end{equation} 
for each Type $j=I,II$ of 2HDM and for each neutral Higgs boson $h=h^{0}$, $% 
H^{0}$, $A^{0}$. While this ratio is not the total branching fraction, it is 
enough for most practical purposes and it is useful in order to compare with 
previous results in the literature. Notice that for $m_{H^{\pm}}>m_{t}$ (the 
most probable situation for Type~II 2HDM's, see below) the ratio (\ref 
{fiducialH}) reduces to $B^{j}(t\rightarrow h+c)=\Gamma^{j}(t\rightarrow 
h+c)/\Gamma(t\rightarrow W^{+}+b)$, which is the one that we used in Ref.  
\cite{NP}. It is understood that $\Gamma^{j}(t\rightarrow h+c)$ above is 
computed from the one-loop diagrams in Fig.~\ref{fig:1}, with all quark 
families summed up in the loop. Therefore, consistency in 
perturbation theory requires to compute $\Gamma(t\rightarrow W^{+}+b)$ and $% 
\Gamma (t\rightarrow H^{+}+b)$ in the denominator of~(\ref{fiducialH}) only 
at the tree-level (for explicit expressions see e.g.~\cite{CGGJS}). As 
mentioned in Sec. 2, we wish to compare our results for the Higgs channels (% 
\ref{Higgschannels}) with those for the gluon channel~(\ref{gchannel}), so 
that we similarly define  
\begin{equation} 
B^{j}(t\rightarrow g+c)=\frac{\Gamma^{j}(t\rightarrow g+c)}{\Gamma 
(t\rightarrow W^{+}+b)+\Gamma^{j}(t\rightarrow H^{+}+b)}\,\,. 
\label{fiducialG} 
\end{equation} 
We have performed a fully-fledged independent analytical and numerical 
calculation of $\Gamma^{j}(t\rightarrow g+c)$ at one-loop in the context of 
2HDM~I and II. Where there is overlapping, we have checked the numerical 
results of Ref.~\cite{GEilam}, but we point out that they agree with us only 
if $\Gamma(t\rightarrow H^{+}+b)$ is included in the denominator of eq.~(\ref 
{fiducialG}), in contrast to what is asserted in that reference in which $% 
B(t\rightarrow g+c)$ is defined without the charged Higgs channel 
contribution. 
 
We have performed part of the analytical calculation of the diagrams for 
both processes~(\ref{Higgschannels}) and~(\ref{gchannel}) by hand and we 
have cross-checked our results with the help of the numeric and algebraic 
programs FeynArts, FormCalc and LoopTools~\cite{FeynArts}, with which we 
have completed the rest of the calculation. In particular, the cancellation 
of UV divergences in the total amplitudes was also verified by hand. In 
addition we have checked explicitly the gauge invariance of the total 
analytical amplitude for the process~(\ref{gchannel}), which is a powerful 
test. And we have confirmed that our code reproduces the SUSY Higgs contribution 
of Ref.\cite{NP} when we turn on the MSSM Higgs mass relations.
 
As mentioned above, a highly relevant parameter is $\tan\beta$, which must 
be restricted to the approximate range  
\begin{equation} 
0.1<\tan\beta\stackm60  \label{tbrange} 
\end{equation} 
in perturbation theory\footnote{% 
Some authors~\cite{Berger} claim that perturbativity allows $\tan\beta$ to 
reach values of order $100$ and beyond, and these are still used in the 
literature. We consider it unrealistic and we shall not choose $\tan\beta$ 
outside the interval~(\ref{tbrange}). Plots versus $\tan\beta$, however, 
will indulge larger values just to exhibit the dramatic enhancements of our 
FCNC top quark rates.}. It is to be expected from the various couplings 
involved in the processes under consideration that the low $\tan\beta$ 
region could be relevant for both the Type~I and Type~II 2HDM's. In 
contrast, the high $\tan\beta$ region is only potentially important for the 
Type~II. However, the eventually relevant regions of parameter space are 
also determined by the value of the mixing angle $\alpha$, as we shall see 
below. 
 
Of course there are several restrictions that must be respected by our 
numerical analysis. Above all the quadratic violations of $SU(2)$ 
``custodial symmetry'' must be within experimental bound. Therefore, the 
one-loop corrections to the $\rho$-parameter from the 2HDM sector cannot 
deviate from the reference SM contribution in more than one per mil 
\thinspace\cite{PDB2000}:  
\begin{equation} 
|\delta\rho^{2HDM}|\leqslant0.001\,.  \label{drho} 
\end{equation} 
From the analytical expression for $\delta\rho$ in the general 2HDM we have 
introduced this numerical condition in our codes. Moreover, non-SUSY charged 
Higgs bosons from Type~II models are subject to a very important indirect 
bound from radiative B-meson decays, specifically the experimental 
measurement by CLEO of the branching fraction $BR(B\rightarrow 
X_{s}\,\gamma) $ -- equivalently $BR(b\rightarrow s\,\gamma)$ at the quark 
level~\cite{CLEO}. At present the data yield:  
\begin{equation} 
BR(b\rightarrow s\,\gamma)=(3.15\pm0.35\pm0.32\pm0.26)\times10^{-4}. 
\label{CLEO} 
\end{equation} 
The charged Higgs contribution to $BR(b\rightarrow s\,\gamma)$ is positive; 
hence the larger is the experimental rate the smaller can be the charged 
Higgs boson mass. From the various analysis in the literature one finds $% 
m_{H^{\pm}}>(165-200)\,\,GeV$ for virtually any $\tan\beta\stackM1$ \cite 
{Borzumati,Chankowski}. This bound does not apply to Type~I models because 
at large $\tan\beta$ the charged Higgs couplings are severely suppressed, 
whereas at low $\tan\beta$ we recover the previous unrestricted situation of 
Type~II models. Therefore, in principle the top quark decay $t\rightarrow 
H^{+}+b$ is still possible in 2HDM~I; but also in 2HDM~II, if $m_{H^{\pm}} $ 
lies near the lowest end of the previous bound, and in this case that decay 
can contribute to the denominator of eqs.~(\ref{fiducialH})-(\ref{fiducialG}% 
). 
 
One may also derive lower bounds to the neutral Higgs masses for these 
models \cite{Krawczyk}. For instance, one may use the Bjorken process $% 
e^{+}e^{-}\rightarrow Z+h^{o}$ and the associated Higgs boson pair 
production $e^{+}e^{-}\rightarrow h^{0}(H^{0})+A^{0}$ to obtain the 
following bounds in most of the parameter space: $m_{h^{0}}+m_{A^{0}}\stackM% 
100\,GeV$ or $\,\stackM150\,GeV$ depending on whether we accept any value of $% 
\tan\beta$ or we impose $\tan\beta>1$ respectively~\cite{OPAL}. 
In each of these cases there is a light mass corner in parameter 
space both in the CP-even and in the CP-odd mass ranges around $% 
m_{h,^{0}A^{0}}=20-30\,\,GeV$~\cite{OPAL}. Notwithstanding, as it is shown by the fit analysis of precision electroweak data in Ref. \cite{Chankowski}, in the 
large $\tan\beta$ region a light $h^{0}$ is statistically correlated with a 
light $H^{\pm}$, so that this situation is not favored by the aforementioned 
bound from $b\rightarrow s\,\gamma$. Moreover, since our interest in Type~II 
models is mainly focused in the large $\tan \beta$ regime, the corner in the 
light CP-even mass range is a bit contrived. At the end of the day  
one finds that, even in the worst situation, the strict experimental limits still allow generic 2HDM neutral scalar bosons as light as $70\,GeV$ or so. As we said, most of 
these limits apply to Type~II 2HDM's, but we will conservatively apply them 
to Type~I models as well. 
 
Finally, for both models we have imposed the condition that the (absolute 
value) of the trilinear Higgs self-couplings do not exceed the maximum 
unitarity limit tolerated for the SM trilinear coupling:  
\begin{equation} 
\ \left| \lambda_{HHH}\right| \leqslant\left| 
\lambda_{HHH}^{(SM)}(m_{H}=1\,TeV)\right| =\frac{3\,g\,(1\,TeV)^{2}}{2\,M_{W}}\,\,. 
\label{unitbound} 
\end{equation} 
 
\begin{figure}[t] 
\begin{center} 
\resizebox{\textwidth}{!}{\includegraphics{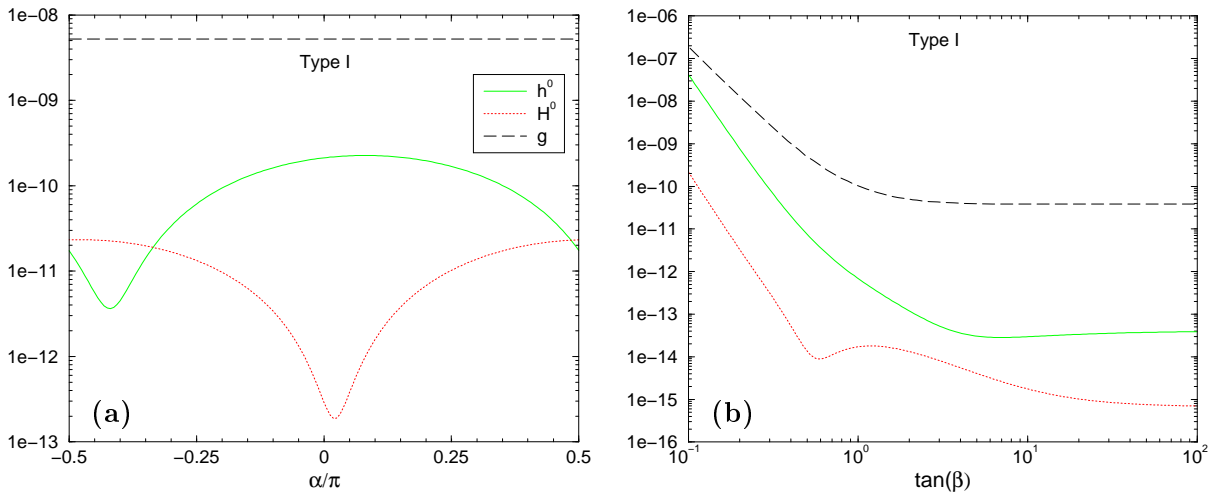}} 
\end{center} 
\caption{Evolution of the FCNC top quark fiducial ratios~(\ref{fiducialH})-(% 
\ref{fiducialG}) in Type~I 2HDM versus: \textbf{(a)} the mixing angle $% 
\protect\alpha$ in the CP-even Higgs sector, in units of $\protect\pi$;  
\textbf{(b)} $\tan\protect\beta$. The values of the fixed parameters are as 
in eqs.~(\ref{inputsmixing}) and (\ref{inputsmasses}).} 
\label{fig:2} 
\end{figure}

\begin{figure}[t] 
\begin{center} 
\resizebox{\textwidth}{!}{\includegraphics{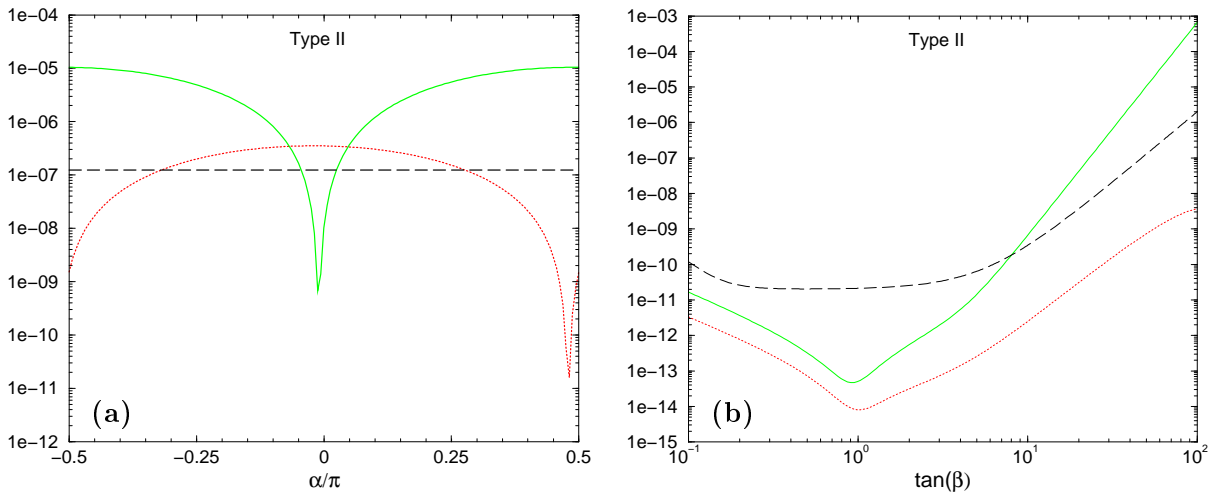}} 
\end{center} 
\caption{As in Fig.~\ref{fig:2}, but for the 2HDM~II. The plot in (b) 
continues above the bound in eq.~(\ref{tbrange}) just to better show the 
general trend.} 
\label{fig:3} 
\end{figure}

\begin{figure}[t] 
\begin{center} 
\resizebox{\textwidth}{!}{\includegraphics{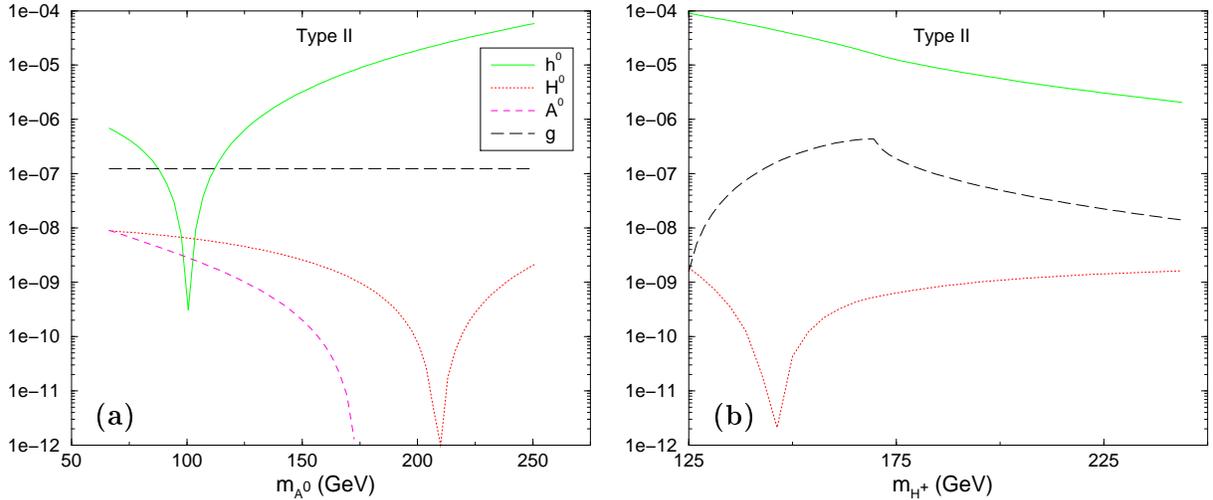}} 
\end{center} 
\caption{Evolution of the FCNC top quark fiducial ratios~(\ref{fiducialH})-(% 
\ref{fiducialG}) in Type~II 2HDM versus: \textbf{(a)} the CP-odd Higgs boson 
mass $m_{A^{0}}$; \textbf{(b) }the charged Higgs boson mass $m_{H^{\pm}}$. 
The values of the fixed parameters are as in eqs.~(\ref{inputsmixing}) and (% 
\ref{inputsmasses}). The plot in (b) starts below the bound $% 
m_{H^{\pm}}>165\,GeV$ mentioned in the text to better show the general 
trend. } 
\label{fig:4} 
\end{figure}

\begin{figure}[t] 
\begin{center} 
\resizebox{\textwidth}{!}{\includegraphics{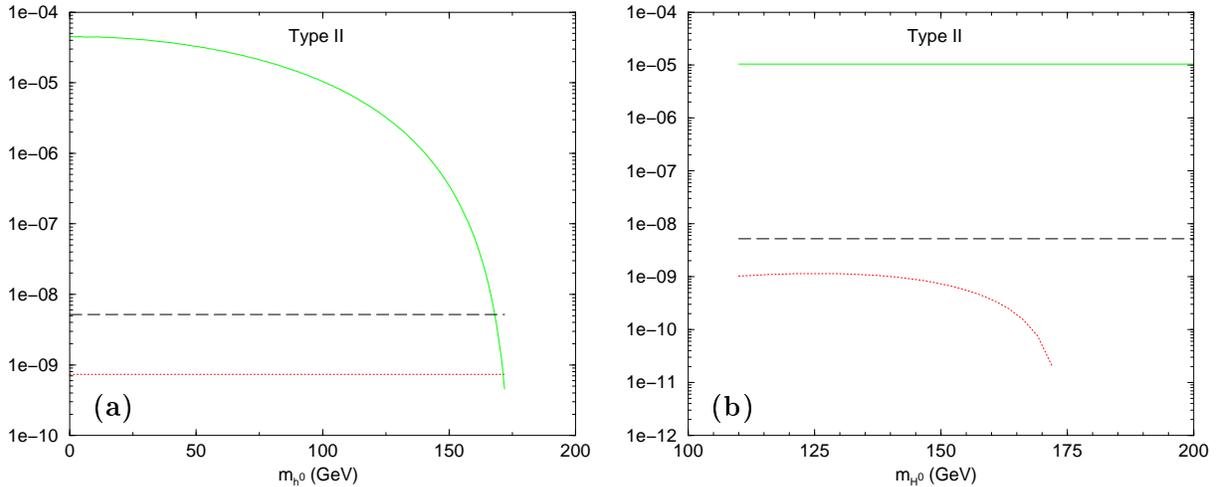}} 
\end{center} 
\caption{As in Fig.~\ref{fig:4}, but plotting versus: \textbf{(a) }the 
lightest CP-even Higgs boson mass $m_{h^{0}}$; \textbf{(b) }the heaviest 
CP-even Higgs boson mass $m_{H^{0}}$.} 
\label{fig:5} 
\end{figure}

The combined set of independent conditions turns out to be quite effective 
in narrowing down the permitted region in the parameter space, as can be 
seen in Figs.~\ref{fig:2}-\ref{fig:5} where we plot the fiducial FCNC rates~(% 
\ref{fiducialH})-(\ref{fiducialG}) versus the parameters~(\ref{freeparam}). 
The cuts in some of these curves just reflect the fact that at least one of 
these conditions is not fulfilled. 
 
After scanning the parameter space, we see in Figs.~\ref{fig:2}-\ref{fig:3} 
that the 2HDM~I (resp. 2HDM~II) prefers low values (resp. high values) of $% 
\tan\alpha$ and $\tan \beta$ for a given channel, e.g. $t\rightarrow 
h^{0}\,c $. Therefore, the following choice of mixing angles will be made to 
optimize the presentation of our numerical results:  
\begin{align} 
\text{2HDM I} & :\ \tan\alpha=\tan\beta=1/4\,\,;  \notag \\ 
\text{2HDM II} & :\ \tan\alpha=\tan\beta=50\,.  \label{inputsmixing} 
\end{align} 
We point out that, for the same values of the masses, one obtains the same 
maximal FCNC rates for the alternative channel $t\rightarrow H^{0}\,c$ 
provided one just substitutes $\alpha\rightarrow\pi/2-\alpha$. Equations~(% 
\ref{inputsmixing}) define the eventually relevant regions of parameter 
space and, as mentioned above, depend on the values of the mixing angles $% 
\alpha$ and $\beta$, namely $\beta \simeq\alpha\simeq0$ for Type~I and $% 
\beta\simeq\alpha\simeq\pi/2$ for Type~II.

Due to the $\alpha \rightarrow \pi /2-\alpha $ symmetry of the maximal rates 
for the CP-even Higgs channels, it is enough to concentrate the numerical 
analysis on one of them, but one has to keep in mind that the other channel 
yields the same rate in another region of parameter space. Whenever a mass 
has to be fixed, we choose conservatively the following values for both 
models:  
\begin{equation} 
m_{h^{0}}=100\,GeV\,,\;m_{H^{0}}=150\,GeV\,,\;\;m_{A^{0}}=m_{H^{\pm 
}}=180\,GeV\,.  \label{inputsmasses} 
\end{equation} 
Also for definiteness, we take the following values for some relevant SM\ 
parameters in our calculation:  
\begin{equation} 
m_{t}=175\,GeV\,,\;m_{b}=5\,GeV\,,\,\,\alpha 
_{s}(m_{t})=0.11\,,\;\, V_{cb}=0.040\,,  \label{SMparam} 
\end{equation} 
and the remaining ones are as in \cite{PDB2000}. Notice that our choice of $% 
m_{A^{0}}$ prevents the decay $t\rightarrow \;A^{0}\,c$ from occurring, and 
this is the reason why it does not appear in Figs.~\ref{fig:2}-~\ref{fig:3}. The variation 
of the results with respect to the masses is studied in Figs.~\ref{fig:4}-% 
\ref{fig:5}. In particular, in Fig.~\ref{fig:4} we can see the (scanty) rate 
of the channel $t\rightarrow A^{0}\,c$ when it is kinematically allowed. In 
general the pseudoscalar channel is the one giving the skimpiest FCNC rate. 
This is easily understood as it is the only one that does not have trilinear 
couplings with the other Higgs particles (Cf. Table~\ref{tab:1}). While it 
does have trilinear couplings involving Goldstone bosons, these are not 
enhanced. The crucial role played by the trilinear Higgs self-couplings in 
our analysis cannot be underestimated as they can be enhanced by playing 
around with both (large or small) $\tan \beta $ \emph{and} also with the 
mass splittings among Higgses. This feature is particularly clear in Fig.~% 
\ref{fig:4}a where the rate of the channel $t\rightarrow h^{0}\,c$ is 
dramatically increased at large $m_{A^{0}}$, for fixed values of the other 
parameters and preserving our list of constraints. Similarly would happen 
for $t\rightarrow H^{0}\,c$ in the corresponding region $\alpha \rightarrow 
\pi /2-\alpha $. 
 
From Figs.~\ref{fig:2}a and \ref{fig:2}b it is pretty clear that the 
possibility to see FCNC decays of the top quark into Type~I Higgs bosons is 
plainly hopeless even in the most favorable regions of parameter space -- 
the lowest (allowed) $\tan\beta$ end. In fact, the highest rates remain 
neatly down $10^{-6}$, and therefore they are (at least) one order of 
magnitude below the threshold sensibility of the best high luminosity top 
quark factory in the foreseeable future (see Section 4). We remark, in Fig.~% 
\ref{fig:2}, that the rate for the reference decay $t\rightarrow g\,c$ in the 
2HDM~I is also too small but remains always above the Higgs boson rates. 
Moreover, for large $\tan\beta$ one has, as expected, $B^{I}\,($ $% 
t\rightarrow g\,c)\rightarrow B^{SM}\,($ $t\rightarrow 
g\,c)\simeq4\times10^{-11}$ because in this limit all of the charged Higgs 
couplings in the 2HDM~I (the only Higgs couplings involved in this decay) 
drop off. Due to the petty numerical yield from Type~I models we refrain 
from showing the dependence of the FCNC rates on the remaining parameters.

Fortunately, the meager situation just described does not replicate for Type 
II Higgs bosons. For, as shown in Figs.~\ref{fig:3}a and \ref{fig:3}b, the 
highest potential rates are of order $10^{-4}$, and so there is hope for 
being visible. In this case the most favorable region of parameter space is 
the high $\tan\beta$ end in eq.~(\ref{tbrange}). Remarkably, there is no 
need of risking values over and around $100$ (which, as mentioned above, are 
sometimes still claimed as perturbative!) to obtain the desired rates. But 
it certainly requires to resort to models whose hallmark is a large value of  
$\tan\beta$ of order or above $m_{t}/m_{b}\stackM35$. As for the dependence 
of the FCNC rates on the various Higgs boson masses (Cf. Figs.~\ref{fig:4}-% 
\ref{fig:5}) we see that for large $m_{A^{0}}$ the decay $t\rightarrow 
h^{0}\,c$ can be greatly enhanced as compared to $t\rightarrow g\,c$; and of 
course, once again, the same happens with $t\rightarrow H^{0}\,c$ in the 
alternative region $\alpha\rightarrow \pi/2-\alpha$. We also note (from the 
combined use of Figs.~\ref{fig:3}b, \ref{fig:4}a and \ref{fig:4}b) that in 
the narrow range where $t\rightarrow H^{+}\,b$ could still be open in the 
2HDM~II, the rate of $t\rightarrow h^{0}\,c$ becomes the more visible the 
larger and larger is $\tan\beta$ and $m_{A^{0}}$. Indeed, in this region one 
may even overshoot the $10^{-4}$ level without exceeding the upper bound~(% 
\ref{tbrange}) while also keeping under control the remaining constraints, 
in particular eq.~(\ref{drho}). Finally, the evolution of the rates~(\ref 
{fiducialH})-(\ref{fiducialG}) with respect to the two CP-even Higgs boson 
masses is shown in Figs.~\ref{fig:5}a and \ref{fig:5}b. The neutral Higgs 
bosons themselves do not circulate in the loops (Cf. Fig.~\ref{fig:1}) but 
do participate in the trilinear couplings (Cf. Table~\ref{tab:1}) and so the 
evolution shown in some of the curves in Fig.~\ref{fig:5} is due to both the 
trilinear couplings and to the phase space exhaustion. 
 
Turning now to the light scalar and pseudoscalar corners in parameter space 
mentioned above, it so happens that, after all, they prove to be of little 
practical interest in our case. Ultimately this is due to the quadratic 
Higgs boson mass differences entering $\delta\rho$ which make very 
difficult to satisfy the bound~(\ref{drho}). The reason being that for Type 
II models the limit $m_{H^{\pm}}\stackM165\,GeV$ from $b\rightarrow 
s\,\gamma $ implies that the constraint~(\ref{drho}) cannot be preserved in 
the presence of light neutral Higgses. In actual fact the analysis shows 
that if e.g. one fixes $m_{h^{0}}=20-30\;GeV$, then the minimum $m_{A^{0}}$ 
allowed by $\delta\rho$ is $100\,GeV$ and the maximum rate~(\ref{fiducialH}) 
is of order $10^{-6}$. Conversely, if one chooses $m_{A^{0}}=20-30\;GeV$, 
then the minimum $m_{h^{0}}$ allowed by $\delta\rho$ is $120\,GeV$ and the 
maximum rate (\ref{fiducialH}) is near $10^{-4}$. Although in the last case 
the maximum rate is higher than in the first case, it is just of the order 
of the maximum rate already obtained outside the light mass corners of 
parameter space. On the other hand, these light mass regions do not help us 
in Type~I models either. Even though for these models we do not have the $% 
b\rightarrow s\,\gamma$ bound on the charged Higgs, we still have the direct 
LEP 200 bound $m_{H^{\pm}}\stackM78.7\,GeV$~\cite{TJunk} which is of course 
weaker than the CLEO bound. As a consequence the $\delta\rho$ constraint can 
be satisfied in the 2HDM~I for neutral Higgs bosons lighter than in the 
corresponding 2HDM~II case, and one does get some enhancement of the FCNC 
rates. Specifically, one may reach up to $10^{-6}$. However, the maximal 
rates~(\ref{fiducialH}) for the 2HDM~I Higgs bosons are so small (see Figs.~% 
\ref{fig:2}a-\ref{fig:2}b) that this order of magnitude enhancement is 
rendered immaterial. The upshot is that the top quark FCNC processes are not 
especially sensitive to the potential existence of a very light Higgs boson 
in either type of 2HDM. 
 
\section{Discussion and conclusions} 
 
The sensitivities to FCNC top quark decays for $100\,fb^{-1}$ of integrated 
luminosity in the relevant colliders are estimated to be~\cite{Frey}:  
\begin{align} 
\mathrm{\mathbf{LHC:}}B(t\rightarrow c\,X) & \gsim5\times10^{-5}\,,  \notag \\  
\mathrm{\mathbf{LC:}}B(t\rightarrow c\,X) & \gsim5\times10^{-4}\,,  
\label{sensitiv} \\ 
\mathrm{\mathbf{TEV33:}}B(t\rightarrow c\,X) & \gsim5\times10^{-3}\,\,\,.  
\notag 
\end{align} 
This estimation has been confirmed by a full signal-background analysis 
for the hadron colliders in Ref.\cite{Saavedra}. 
From these experimental expectations and our numerical results it becomes 
patent that whilst the Tevatron will remain essentially blind to this kind 
of physics, the LHC and the LC will have a significant potential to observe 
FCNC decays of the top quark beyond the SM. Above all there is a possibility 
to pin down top quark decays into neutral Higgs particles, eq.~(\ref 
{Higgschannels}), within the framework of the general 2HDM~II provided $% 
\tan\beta\stackM\,m_{t}/m_{b}\sim35$. The maximum rates are of order $% 
10^{-4} $ and correspond to the two CP-even scalars. This conclusion is 
remarkable from the practical (quantitative) point of view, and also 
qualitatively because the top quark decay into the SM Higgs particle is, in 
notorious contradistinction to the 2HDM~II case, the less favorable top 
quark FCNC rate in the SM. On the other hand, we deem practically hopeless 
to see FCNC decays of the top quark in a general 2HDM~I for which the 
maximum rates are of order $10^{-7}$. This order of magnitude cannot be 
enhanced unless one allows $\tan\beta\ll0.1$, but the latter possibility is 
unrealistic because perturbation theory breaks down and therefore one cannot 
make any prediction within our approach. 
 
We have made a parallel numerical analysis of the gluon channel $% 
t\rightarrow c\;g$ in both types of 2HDM's. We confirm that this is another 
potentially important FCNC mode of the top quark in 2HDM extensions of the 
SM~\cite{GEilam} but, unfortunately, it still falls a bit too short to be 
detectable. The maximum rates for this channel lie below $10^{-6}$ in the 
2HDM~I (for $\tan\beta>0.1)$ and in the 2HDM~II (for $\tan\beta<60$), and so 
it will be hard to deal with it even at the LHC. 
 
We are thus led to the conclusion that the Higgs channels~(\ref{Higgschannels}), 
more specifically the CP-even ones, give the highest potential rates for top 
quark FCNC decays in a general 2HDM~II. Most significant of all: they are  
\emph{the only} FCNC decay modes of the top quark, within the simplest 
renormalizable extensions of the SM, that have a real chance to be seen in 
the next generation of high energy, high luminosity, colliders. 
 
The former conclusions are similar to the ones derived in Ref.~\cite{NP} for 
the MSSM case, but there are some conspicuous differences on which we wish 
to elaborate a bit in what follows~\cite{Carmel}. First, in the general 
2HDM~II the two channels $t\rightarrow(h^{0},H^{0})\;c$ give the same 
maximum rates, provided we look at different (disjoint) regions of the 
parameter space. The $t\rightarrow A^{0}\;c$ channel is, as mentioned, 
negligible with respect to the CP-even modes. Hereafter we will discard this 
FCNC top quark decay mode from our discussions within the 2HDM context. On 
the other hand, in the MSSM there is a most distinguished channel, viz. $% 
t\rightarrow h^{0}\;c$, which can be high-powered by the SUSY stuff all over 
the parameter space. In this framework the mixing angle $\alpha$ becomes 
stuck once $\tan\beta$ and the rest of the independent parameters are given, 
and so there is no possibility to reconvert the couplings between $h^{0}$ 
and $H^{0}$ as in the 2HDM. Still, we must emphasize that in the MSSM the 
other two decays $t\rightarrow H^{0}\;c$ and $t\rightarrow A^{0}\;c$ can be 
competitive with $t\rightarrow h^{0}\;c$ in certain portions of parameter 
space. For example, $t\rightarrow H^{0}\;c$ becomes competitive when the 
pseudoscalar mass is in the range $110\,GeV<m_{A^{0}}\stackm170\,GeV$~\cite 
{NP}. The possibility of having more than one FCNC decay~(\ref{Higgschannels}% 
) near the visible level is a feature which is virtually impossible in the 
2HDM~II. Second, the reason why $t\rightarrow h^{0}\;c$ in the MSSM is so 
especial is that it is the only FCNC top quark decay~(\ref{Higgschannels}) 
which is always kinematically open throughout the whole MSSM parameter 
space, while in the 2HDM all of the decays (\ref{Higgschannels}) could be, 
in the worse possible situation, dead closed. Nevertheless, this is not the 
most likely situation in view of the fact that all hints from high precision 
electroweak data seem to recommend the existence of (at least) one 
relatively light Higgs boson~\cite{TJunk,Hagiwara}. This is certainly an 
additional motivation for our work, as it leads us to believe that in all 
possible (renormalizable) frameworks beyond the SM, and not only in SUSY, we 
should expect that at least one FCNC decay channel (\ref{Higgschannels}) 
could be accessible. Third, the main origin of the maximum FCNC rates in the 
MSSM traces back to the tree-level FCNC couplings of the gluino~\cite{NP}. 
These are strong couplings, and moreover they are very weakly restrained by 
experiment. In the absence of such gluino couplings, or perhaps by further 
experimental constraining of them in the future, the FCNC rates in the MSSM 
would boil down to just the electroweak (EW) contributions, to wit, those 
induced by charginos, squarks and also from SUSY Higgses. The associated 
SUSY-EW rate is of order $10^{-6}$ at most, and therefore it is barely 
visible, most likely hopeless even for the LHC. In contrast, in the general 
2HDM the origin of the contributions is purely EW and the maximum rates are 
two orders of magnitude higher than the full SUSY-EW effects in the MSSM. It 
means that we could find ourselves in the following situation. Suppose that 
the FCNC couplings of the gluino get severely restrained in the future and 
that we come to observe a few FCNC decays of the top quark into 
Higgs bosons, perhaps at the LHC and/or the LC. Then we would immediately 
conclude that these Higgs bosons could not be SUSY-MSSM, whilst 
they could perhaps be CP-even members of a 2HDM~II. Fourth, the gluino 
effects are basically insensitive to $\tan\beta$, implying that the maximum 
MSSM rates are achieved equally well for low, intermediate or high values of  
$\tan\beta,$ whereas the maximum 2HDM~II rates (comparable to the MSSM ones) 
are attained only for high $\tan\beta$. 
 
The last point brings about the following question: what could we possibly 
conclude if the gluino FCNC couplings were not further restricted by 
experiment and the tagging of certain FCNC decays of the top quark into 
Higgs bosons would come into effect? Would still be possibly to discern 
whether the Higgs bosons are supersymmetric or not? The answer is, most 
likely yes, provided certain additional conditions would be met. 
 
There are many possibilities and corresponding strategies, but we will limit 
ourselves to point out some of them. For example, let us consider the type 
of signatures involved in the tagging of the Higgs channels. In the favorite 
FCNC region~(\ref{inputsmixing}) of the 2HDM~II, the combined decay  $t\rightarrow 
h\;\,c\rightarrow cb\overline{b}$ is possible only for $h^{0}$ or for $H^{0}$% 
, but not for both -- Cf. Fig.~\ref{fig:3}a -- whereas in the MSSM, $h^{0}$ 
together with $H^{0}$, are highlighted for $110\,GeV<m_{A^{0}}<m_{t}$, with 
no preferred $\tan \beta $ value. And similarly, $t\rightarrow A^{0}\;c$ is 
also non-negligible for $m_{A^{0}}\stackm120\,GeV$\thinspace \cite{NP}. Then 
the process $t\rightarrow h\;\,c\rightarrow cb\overline{b}$ gives 
rise to high $p_{T}$ charm-quark jets and a recoiling $b\overline{b}$ pair 
with large invariant mass. It follows that if more than one distinctive 
signature of this kind would be observed, the origin of the hypothetical 
Higgs particles could not probably be traced back to a 2HDM~II.  
 
One might worry that in the case of $h^{0}$ and $H^{0}$ they could also (in 
principle) decay into electroweak gauge boson pairs $h^{0},H^{0}\rightarrow 
V_{ew}\overline{V}_{ew}$, which in some cases could be kinematically 
possible. But this is not so in practice for the 2HDM~II if we stick to our 
favorite scenario, eq.~(\ref{inputsmixing}). In fact, we recall that the 
decay $h^{0}\rightarrow V_{ew}\overline{V}_{ew}$ is not depressed with 
respect to the SM Higgs boson case provided $\beta -\alpha =\pi /2$, and 
similarly for $H^{0}\rightarrow V_{ew}\overline{V}_{ew}$ if $\beta -\alpha =0 
$. However, neither of these situations is really pinpointed by FCNC physics 
because we have found $\beta \simeq \pi /2$ in the most favorable region of 
our numerical analysis, and moreover $\alpha $ was also seen there to be 
either $\alpha \simeq \pi /2$ (for $h^{0}$) or $0$ (for $H^{0}$), so both 
decays $h^{0},H^{0}\rightarrow V_{ew}\overline{V}_{ew}$ are suppressed in 
the regions where the FCNC rates of the parent decays $t\rightarrow 
(h^{0},H^{0})\;\,c$ are maximized. Again, at variance with this situation, 
in the MSSM case $H^{0}\rightarrow V_{ew}\overline{V}_{ew}$ is perfectly 
possible -- not so $h^{0}\rightarrow V_{ew}\overline{V}_{ew}$ due to the 
aforementioned upper bound on $m_{h^{0}}$ -- because $\tan \beta $ has no 
preferred value in the most favorable MSSM decay region of $t\rightarrow 
H^{0}\;\,c$. Therefore, detection of a high $p_{T}$ charm-quark jet against 
a $V_{ew}\overline{V}_{ew}$ pair of large invariant mass could only be 
advantageous in the MSSM, not in the 2HDM. Similarly, for $\tan \beta \stackM% 
1$ the decay $H^{0}\rightarrow h^{0}\;h^{0}$ (with real or virtual $h^{0}$) 
is competitive in the MSSM~\cite{Djouadi} in a region where the parent FCNC 
top quark decay is also sizeable. Again this is impossible in the 2HDM~II 
and therefore it can be used to distinguish the two (SUSY and non-SUSY) 
Higgs frames. 
 
Finally, even if we place ourselves in the high $\tan\beta$ region both for 
the MSSM and the 2HDM~II, then the two frameworks could still possibly be 
separated provided that two Higgs masses were known, perhaps one or both of 
them being determined from the tagged Higgs decays themselves, eq.~(\ref 
{Higgschannels}). Suppose that $\tan\beta$ is numerically known (from other 
processes or from some favorable fit to precision data), then the full 
spectrum of MSSM Higgs bosons would be approximately determined (at the tree 
level) by only knowing one Higgs mass, a fact that could be used to check 
whether the other measured Higgs mass becomes correctly predicted. Of 
course, the radiative corrections to the MSSM Higgs mass relations can be 
important at high $\tan\beta$~\cite{Hollik}, but these could be taken into 
account from the approximate knowledge of the relevant sparticle masses 
obtained from the best fits available to the precision measurements within 
the MSSM. If there were significant departures between the predicted mass 
for the other Higgs and the measured one, we would probably suspect that the 
tagged FCNC decays into Higgs bosons should correspond to a 
non-supersymmetric 2HDM~II. 
 
At the end of the day we see that even though the maximum FCNC rates for the 
MSSM and the 2HDM~II are both of order $10^{-4}$ -- and therefore 
potentially visible -- at some point on the road it should be possible to 
disentangle the nature of the Higgs model behind the FCNC decays of the top 
quark. Needless to say, if all the recent fuss at CERN~\cite{TJunk} about 
the possible detection of a Higgs boson would eventually be confirmed, this 
could still be interpreted as the discovery of one neutral member of an 
extended Higgs model. Obviously the combined Higgs data from LEP 200 and the 
possible discovery of FCNC top quark decays into Higgs bosons at the LHC/LC 
would be an invaluable cross-check of the purportedly new phenomenology. 
 
We emphasize our most essential conclusions in a nutshell: i) Detection of 
FCNC top quark decay channels into a neutral Higgs boson would be a blazing 
signal of physics beyond the SM; ii) There is a real chance for seeing rare 
events of that sort both in generic Type~II 2HDM's and in the MSSM. The 
maximum rates for the leading FCNC processes~(\ref{Higgschannels}) and (\ref 
{gchannel}) in the 2HDM~II (resp. in the MSSM) satisfy the relations  
\begin{equation} 
BR(t\rightarrow g\,c)<10^{-6}(10^{-5})<BR(t\rightarrow h\,c)\sim10^{-4}\,, 
\label{summary} 
\end{equation} 
where it is understood that $h$ is $h^{0}$ or $H^{0}$, but not both, in the 
2HDM~II; whereas $h$ is most likely $h^{0}$, but it could also be $H^{0}$ 
and $A^{0}$, in the MSSM ; iii) Detection of more than one Higgs channel 
would greatly help to unravel the type of underlying Higgs model. 
 
The pathway to seeing new physics through FCNC decays of the top quark is 
thus potentially open. It is now an experimental challenge to accomplish 
this program using the high luminosity super-colliders round the corner. 
 
\bigskip 
 
\vskip5mm \noindent\textbf{Acknowledgments.} One of the authors (J.S) would 
like to thank H. Haber and T. Junk for helpful discussions. The work of S.B. 
and J.S. has been supported in part by CICYT under project No. AEN99-0766. 
The work of J.G. has been supported by the European Union under contract No. 
HPMF-CT-1999-00150.  

%%%%%%%%%%%%%%%%%%%%%%%%%%%%%%%%%%%%%%%%%%%%%%%%%%%%%%%   
%  
%  
%  
%  
%  
%  

\end{document}